*Article*

# Systematic Review on Reinforcement Learning in the field of Fintech

Nadeem Malibari [1,†,‡] 0000-0003-3702-3743 , Iyad Katib [1,‡] and Rashid Mehmood [2,*]

1 Department of Computer Science, Faculty of Computing and Information Technology;King Abdulaziz University, Jeddah 21589, Saudi Arabia; nmahimalibari@stu.kau.edu.sa;IAKatib@kau.edu.sa
2 High Performance Computing Center, King Abdulaziz University, Jeddah 21589, Saudi Arabia;RMehmood@kau.edu.sa
‡ These authors contributed equally to this work.

**Abstract:** Applications of Reinforcement Learning in the Finance Technology (Fintech) have acquired a lot of admiration lately. Undoubtedly Reinforcement Learning, through its vast competence and proficiency, has aided remarkable results in the field of Fintech. The objective of this systematic survey is to perform an exploratory study on a correlation between reinforcement learning and Fintech to highlight the prediction accuracy, complexity, scalability, risks, profitability and performance. Major uses of reinforcement learning in finance or Fintech include portfolio optimization, credit risk reduction, investment capital management, profit maximization, effective recommendation systems, and better price setting strategies. Several studies have addressed the actual contribution of reinforcement learning to the performance of financial institutions. The latest studies included in this survey are publications from 2018 onward. The survey is conducted using PRISMA technique which focuses on the reporting of reviews and is based on a checklist and four-phase flow diagram. The conducted survey indicates that the performance of RL-based strategies in Fintech fields proves to perform considerably better than other state-of-the-art algorithms. The present work discusses the use of reinforcement learning algorithms in diverse decision-making challenges in Fintech and concludes that the organizations dealing with finance can benefit greatly from Robo-advising, smart order channelling, market making, hedging and options pricing, portfolio optimization, and optimal execution.

## 1. Introduction

The contemporary Financial technology (Fintech) theory has fast acquired popularity both as an academic research framework and as a practical application in the financial services sector. The volume of data has grown significantly due to presence of a range of variables, such as transactions, quotations, trading, and order flow, thereby posing new computational and theoretical issues. Fintech, being a blend of the two concepts (finance and technology), is a rising academic and professional area that tries to use automation and other technologies to simplify and improve the provision of financial services. The portmanteau term "Fintech" refers to the use of high-tech computer software and algorithmic procedures to enhance the financial management capabilities of professionals and businesses. The conventional use of desktop computers is increasingly being augmented with mobile devices, including tablets and cellphones. The objective of any investment in the financial market is to maximize profits while minimizing risks. Artificial intelligence (AI) is largely utilized to address challenges that seem to demand high levels of intellect and are difficult for people to handle. Expert systems may be used to assist people in making choices, as rule-based systems are often utilized in this context. Manual modeling of knowledge bases of such systems involves a variety of obstacles that requires a large amount of work, and is prone to errors. Some people refer to this problem as the "knowledge acquisition bottleneck" [1].

The bottleneck in the knowledge acquisition process can be handled by those systems which are capable of giving more accurate and quick responses while being error-free to handle the challenges of large data sets. Machine learning-based knowledge has been used in a broad variety of domains, including health, finance, behavior analysis, natural language processing (NLP), and other areas [2–4]. In the financial market, decisions on purchase and sell trading may be made either by humans or by computer intelligence. Over the course of the last several decades, the use of computers to make trading decisions while participating in the foreign exchange and stock markets, has progressively risen. In addition, the fields that are directly associated with AI and machine learning, are used in a wide variety of other domains as well.

The objective of Reinforcement Learning (RL), a subfield of computer science which falls under the umbrella of AI, is to teach computers to solve problems on their own without being given specific instructions to do so [1]. Instead of relying on more typical stochastic control systems, it has been chosen to build new ideas that are based on RL so that all of this information may be put into use. RL is a trial-and-error approach to train the computers to produce autonomous agents with optimal interactions with their environments. The theory of RL describes how agents may learn to take optimum and more accurate behaviors via iterative trial and error. This may be expressed more formally as the goal of all agents (whether they are humans, animals, or technology) to maximize the discounted value of future rewards over time [5].

The notion of RL has not only been embraced by the Fintech industry, but it has also been employed by a wide variety of other businesses. RL is a kind of machine learning and training based on the principle of rewarding favorable behaviors and discouraging the bad ones, as defined by [6][7]. For the most part, a RL agent can "scan" its surroundings, "think" about what to do next, and "learn" from its mistakes. The RL approach, thus, provides high rewards for agents who demonstrate excellent behavior as well as punishments for those who show poor behavior. As a result of the incentive structure, the agent is motivated to find a solution that is beneficial to both the current and future parties in order to reach the best possible outcome [8]. Long-term objectives serve as a shield against the agent's tendency to focus on more immediate needs. As a result, the agent eventually learns how to avoid potentially harmful situations by seeking out individuals who have a good attitude and can produce compatible offspring. AI is quickly gaining ground in the area of RL with punishments and rewards, due to its effectiveness as a method of controlling unsupervised learning.

RL encompasses several algorithms, instead of referring to a specific algorithm, each of which takes a unique approach to the problem being addressed. The primary reason for the disparities between the algorithms is the approach that they use while attempting to investigate and exploit their surroundings. State-action-reward-state-action (SARSA) algorithm resembles to Q-learning and Deep Q-Networks (DQN); the three are the most frequently used algorithms in deep reinforcement learning (DRL). The first step in SARSA's functioning is to appoint a policy to the agent. The agent is given the opportunity to learn about the chance that certain behaviors, in addition to receiving rewards, may yield beneficial states via the use of the policy. Q-learning adopts the complete opposite approach as compared to SARSA. To be more specific, the agents are not provided with any kind of policy, which emphasizes the fact that their investigation of the surrounding world is self-directed. On the other hand, DQN in RL (Figure 1) make use of neural networks in conjunction with other RL methods. The RL strategy of self-directed ecosystem exploration is often used by these algorithms. Therefore, future actions are based on a sample population of past behaviors that were advantageous and were acquired by the neural network.

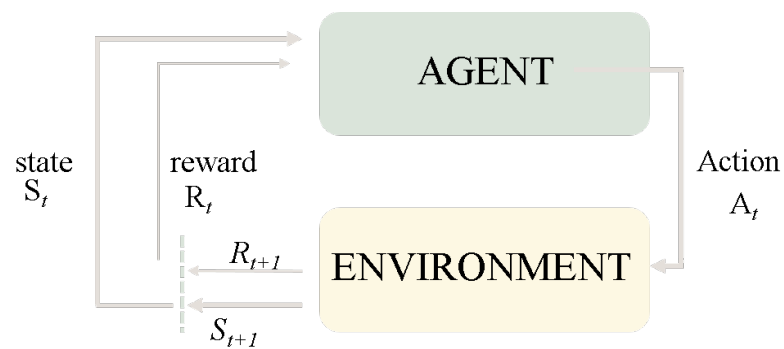

**Figure 1.** Reinforcement learning architecture

Figure 1 illustrates the actual mechanism in which the RL approach functions; clarifying all the agents, elements and concepts entailed in the RL design. Compared to supervised and unsupervised learning environments, RL is quite different regarding the overall purpose [9]. Precisely, while the main target of unsupervised learning is to determine the differences and similarities between data points, the overall goal of RL is to identify an appropriate action design that can optimize the aggregate agent reward [7].

RL is widely used in the Fintech sector for a variety of purposes. One of the most widespread implementations of RL in the financial sector is the use of trading bots, which are programmed to learn from the stock and trading market environment via interactions with the actual market [10]. These RL bots use trial and error to fine-tune their learning processes according to the unique characteristics of each commodity traded on the stock market. These methods are based on the specifics of each stock. In order to foster the application of DRL in Fintech and understand the dynamics of the financial business, systematic literature surveys on this topic are necessary. This survey reassures that the introduction of RL tools has a profound impact on the global financial sector.

Another new field of research is the use of RL to anticipate stock prices, alongside trading bots. RL for stock price prediction has seen far less attention than the supervised, and at times, unsupervised machine learning algorithms, despite the fact that it has the potential to provide surprisingly accurate outcomes. Some problems have been discovered with supervised learning-based stock price forecasts; for example, due to its chaotic and unpredictable nature, stock trading is often considered one of the most challenging applications. By including a market volatility component into reward functions, we may be able to increase the size of the trade shares that have low volatility, and vice versa [11]. Issues concerning the supervised learning-based stock price predictions have been identified; one of those issues is that they are not capable of dealing with time-delayed rewards adequately [12]. This can be described alternatively that the delayed penalty or reward is not the foremost concern of supervised learning-based algorithms; rather they only focus on the accuracy of the price prediction at that moment. In addition to that, whereas most supervised machine learning algorithms can only prescribe actions for specific stocks, RL can get us to the decision-making stage right away, allowing us to choose whether to purchase, hold, or sell any stock.

Further applications of RL may be found in other area of Fintech, especially portfolio management and optimization operations. The initial stages in the process of managing a customer's portfolio are to produce an investment policy statement and to find out the criteria specified by the client. Following these phases, the succeeding ones consist of developing a portfolio, monitoring and re-balancing that portfolio, doing security research, allocating assets, evaluating performance, and lastly reporting on the process. The correlation between automated trading and excessive volatility has been highlighted by researchers. When the price of a stock falls below a certain level, automated trading systems have the potential to carry out the pre-programmed rules included inside their trading algorithms. As these algorithms make their decisions based on historical data, they may be more sensitive to changes in price. As a result, a particular strategy may enable investors to quickly optimize and re-balance their portfolios by entering and exiting the market in response to sudden fluctuations in price [8].

Furthermore, RL algorithms have shown good performance, despite the fact that under specific situations they might be vulnerable to certain types of attacks. RL has a small number of applications and may be difficult to put into practice, although it has a significant amount of unrealized potential. The fact that this kind of machine learning is dependent on environment exploration is one of the obstacles that must be addressed before it can be used [13]. The situation is analogous to the case when a robot depended on RL is employed to navigate around a difficult physical environment, for example, the robot would look for new states and pick up new behaviors. Nevertheless, because of the speed and the volatility with which the financial environment changes in the real world, it might be difficult to make the proper judgments on a constant basis [6]. Due to the amount of time required to guarantee that the learning is carried out in an effective manner, the method's demand on computational resources may be substantial, even it its trivial nature may place certain constraints on its application. The complexity of the training environment results in increased amount of time and computational resources required. When there is an adequate quantity of data available, supervised learning has the potential to provide better outcomes more quickly and more effectively than what RL does. This is because it requires less resources than RL to put into practice.

A broad application of our research is demonstrated by its academic significance as well as its non-technical understanding in the context of RL theories for Fintech. The main objectives of this study are outlined as follows:

- Providing in-depth insights on the framework of several RL ideas in the Fintech industry.
- Presenting a non-technical understanding of how RL might improve the process of evaluating financial risk, as well as highlighting the primary benefits and disadvantages of RL-based trading and optimizations.
- Proposing a taxonomy and a generic architecture for the use of RL in financial technology.
- Outlining the most significant drawbacks and difficulties associated with RL in the Fintech industry.

The role of AI and RL in the Fintech sector is discussed in section 2. Methodology to conduct the current survey based on PRISMA technique is presented in Section 3. Section 4 presents taxonomy, as a classification scheme of this study. The Section 5 covers a review of the RL applications in different sectors as highlighted by the studies included in this survey. In the section 6, our in-depth findings are explored and speculated on future directions of our work. The conclusion is drawn in Section 7.

## 2. Role of Reinforcement Learning in the FinTech Sector

In the recent years, significant contribution has been made in the field of machine learning to achieve a certain level of accuracy within limited time constraints. The research work presented by Serrano [14] showed a unique learning algorithm with the Random Neural Network based on the genome model. The work replaced and modified the genetics algorithm where the neuron weights, instead of the neurons themselves, were passed on to the future generations. The idea was to imitate the human brain using a complex deep learning structure. The current decisions were made locally in a fast manner with the help of RL. The identity and memory are provided by deep learning clusters. The ultimate strategic decisions are made by deep learning management clusters, and knowledge is passed down to the future generations via genetic learning. The suggested algorithm and Deep Learning framework were tested in a Fintech Model, which was a Smart Investment application. The Intelligent Bankers in this model are skilled at determining which assets, markets, and risks are advantageous to acquire and sell. The obtained results were promising. The proposed model, which was based on the random neural network model and genetics algorithm along with deep learning, showed that AI is similar to biological learning in the sense that both learn gradually and continuously while adjusting to its environment.

Similarly, the findings of the study performed by Mousavi *et al.* [15] outlined the complete overview of DRL regarding its applications in Fintech. The work provided one of the best background information on RL as the authors clarified the core components of the technology, including exploration, planning, model, reward, policy, and DQNN (Deep Q Neural Network). The comprehensive overview further discusses critical mechanisms of RL, including hierarchical RL, multi-agent RL, transfer learning, unsupervised learning, memory and attention [16].

The work of Bazarbash [17] also addresses the inclusion of Fintech in finance, with machine learning being applied in appraising credit risk. Their work incorporated potential strengths and fragility in the credit assessment using machine learning-based techniques and strategies. According to the author(s), recent developments in digital technology, as well as big data, have enabled Fintech lending to serve as a potentially feasible alternative aimed at lessening credit costs for financial trading. However, relevance of data needs to be ensured due to its vital and central role in machine learning-based analysis. On the other hand, the variables that are responsible for activating and triggering prejudice should not be used for the sole purpose of avoiding digital financial debarring and exclusion. They concluded with the fact that for the accessibility of Fintech credit, there should be a development of technological infrastructure for big data-based decision making. In addition, it was concluded that there should be constant reviewing of machine learning models by analysts for avoiding potential weaknesses in credit rating. Finally, speaking of machine learning-based techniques in the finance sector have proven their worth and they are ready to be deployed.

Hu and Lin [18] examined DRL (Deep RL) to maximize the performance of the financial sector, as it has performed good in several other computational areas. They examined how DRL can be adopted in maximizing portfolio management in the financial sector. The scholars contend that DRL is one of the emerging AI research topics that integrate deep learning for RL and policy optimization for goal-adjusted self-learning without involving any form of human intervention. The idea of their project revolved around the exploration of deep recurrent neural network models for the decision-making of measures on policy optimization in non-Markov decision processes. They evaluated total rewards for policy by crafting a feasible and practicable risk-adjusted reward function. Additionally, the study continued with the entitlement and authorization of combining RL and deep learning with the goal of boosting their already existing capabilities and capacities to discover a paradigmatic and standard policy. Lastly, RL approaches were inquired about and scrutinized to be integrated with deep learning approaches with the impetus of puzzling out the policy optimization query. Despite the deep learning-based approaches, a fair amount of work has been done using RL-based techniques for portfolio optimization.

In regards to the prime task of Portfolio optimization, a study showed a framework of a RL algorithm for order execution and portfolio management. Wang *et al.* [19] addressed the shortcomings of the existing impracticalities and introduced the hierarchical system that is reinforced in stock trading for portfolio management and optimization. These studies were implemented to be viewed for results in the markets of US (United States) and China. The substantial experimental results proved that Hierarchical Reinforced Trading System for Portfolio Management (HRPM) was a notable improvement with regards to other approaches and techniques. HRPM supports and simplifies decision-making tasks which is the prime need for a successful journey in finance fields.

Fintech agents and some of the established financial sector firms have gotten stronger over the pandemic [6]. Although many financial institutions have been affected; even still, there are many more that are fast changing to provide financial services that are tailored for the modern environment [11]. Well before the the onset of Covid-19, certain financial services organizations were already bolstering business models presented by them with cutting-edge and creative Hi-Tech solutions. Today, this procedure has been accelerated even more. The finance sector, in particular, is changing several processes through the use of AI and machine learning.

Hambly *et al.* [20] aimed at surveying the latest growth and expansion in the use of RL approaches and techniques in finance. They begin with providing a detailed introduction to the Markov decision processes followed by various other algorithms. Among these algorithms, their main focus was value and policy-based techniques as these techniques proved to be assumption-free. They concluded with an in-depth discussion on the future technology being supervised by RL. They found several research studies in literature that have addressed (and continue to address) the place of RL in Fintech. The rapid shifts in the finance sector because of the ever-rising volume of data have completely transformed the methods utilized in data analysis and processing. The shifts have also introduced new computational and theoretical challenges. Joseph M. Carew [7] also noted that compared to conventional stochastic control theory as well as numerous other analytical methods for solving fiscal decision-making issues that heavily depend on model assumptions, developments in RL can entirely rely on the large volumes of financial data. Accordingly, these developments can enhance decision support systems in complex financial ecosystems.

Remarkable work in the use of RL as a base to propose high-efficiency models has been reported in literature. RL being the base is paving the way for ground-breaking discoveries in the field of Fintech which is highly appreciable. Kuo *et al.* [8] formed the basis of improvement in trading via RL by proposing a general adversarial market model. He studied that portfolio management was being handled extremely extraordinarily by RL yet some limitations were there that needed to be addressed and resolved. The experimentation included the formation of a virtual market via a market behaviour simulator and a pragmatic security matching system. This virtual market was being used as a front face for the interactive training environment in RL-based portfolio agents. The results showed a 4% improvement in portfolio performances over the state-of-the-art.

As transformation processes in the digital world have changed globally, innumerable financial cyber crimes have also sprouted. The actual silver lining to this is the fact that machine learning and AI have allowed users and companies to safeguard their accounts from fraud and any other forms of malicious usage [21]. Technologies such as block-chain and crypto-currencies have often been closely linked with financial cyber-security. Hendershott *et al.* [21] presented an overview of various issues in Fintech research with an anticipation that their work could stimulate future research in the field of Fintech bu utilizing AI and block-chain correlation. In future, however, digital security will be associated more with AI and machine learning to prevent the existing forms of money laundering [14]. This is because algorithms can adequately and effectively detect as well as notify users of malicious activities. RL as technology will continue to monitor unusual activities or trends; thus, limiting the need for physical vigilance.

Similarly, Cao *et al.* [22] reviewed the research in AI and Fintech over time. Their findings involve many latest updates and upgrades made for banking, trading, training, block-chain and crypto-currencies. Additionally, it makes use of data science and AI methods including data analytics, deep learning, federated learning, privacy-preserving processing, augmentation, optimization, and system intelligence for improvements. Some other strategies include quantitative approaches, complex system methods, intelligent interactions, recognition and responses, and intelligent interactions. The authors provide a very thorough research summary of each of these strategies that make it possible for new disruptive technologies to open up a wide range of opportunities for both users and businesses. In practice people believe that AI and machine learning are just appropriate for huge businesses with large pools of finance and tech professionals. However, the truth can never be further from reality [6] because these technologies, along with potent apps, are being used by Fintech organizations of all sizes for a variety of objectives.

A study found by Tian *et al.* [23] discussed and reviewed in detail the data-driven approaches in Fintech. The research work attempted on providing an extensive comparison inclusive of the advantages and disadvantages of various data-driven algorithms in applications related to finance. Also this work aimed at forming a firm foundation for future discoveries of data-driven approaches in the Fintech domain. The authors present an exploration of data-driven approaches and machine learning algorithms in the fields of portfolio and risk management, sentiment analysis and data privacy protection etc. The existing Finech projects are compared using both conventional data analytics methods and cutting-edge innovations. The framework for the analytical process is created, and in this area, insights are given on implementation, regulation, and workforce development.

Over time, several inflexible and predictive approaches have been suggested to indicate stock price movements, especially the shifts that have failed to obtain satisfactory outcomes [21]. This is specific to when there are issues or crashes in the stock market. Yang *et al.* [24] outlined that to cope with the emerging issues, prediction frameworks have been proposed using deep learning, RL, and adversarial training approaches. RL and field experiments were aggregated. They viewed that by automating two jobs, such as identifying market conditions and executing trading strategies, AI approaches can assist quantitative trading. The representation of high-frequency financial data and striking a balance between the exploitation and exploration of the trading agent with AI approaches are two hurdles that the current methods in quantitative trading must overcome. The main focus of the study was loan debt collection with regards to following a strict sequential collection strategy rather than following private information-based actions. The results suggested the use of fewer collection actions with more vigilance by the loan tenets. Further, the significance of personalization in debt collections was revealed by the results of borrower profiling analysis. This study adds to the body of literature about AI in Fintech by offering tangible, workable and thrifty policy implications.

Khuwaja *et al.* [25] examined adversarial and deep learning networks for applications of Fintech by utilizing heterogeneous information origins. The study argues that the dynamic aspect and increasing sophistication are the principal provocations for Fintech modelling applications like the stock market. The proposal begins with modifications in the newton-divided difference polynomial (NDDP) for the attribution of data that is missing. Long and short term memory network (LSTM) was used for extracting the native and innate properties of the financial market.

The two adversarial networks studied by Khuwaja *et al.* [25] were the confrontational Q learning and HDRM Q learning, with HDFM being heterogeneous data fusion representing market cast. The training of the networks was started and the sole purpose was to improve the prediction level considering the times when the financial market will be evaporative and volatile. The result shown that the performance of prediction was way more improved and upgraded by using the global indicators and proposed adversarial network in comparison to the already existing works and networks.

A detailed survey performed by Cai *et al.* [13] on DRL for data analytics paved way for many other pieces of research and progresses. The survey begins with the vast and foremost introduction to DRL followed by theories and key concepts of DRL. It continues with the discussion of the deployment of DRL on database systems which is the subject of the following section. This facilitates data processing and analytics in a variety of ways, inclusive of data organizing, scheduling, tuning, and indexing. The use of DRL in the analytics and processing of the data is then surveyed, covering everything from data preparation and NLP to healthcare and Fintech. Finally, they went over several significant unresolved issues and potential future research areas for DRL in the analytics and processing of the data.

Although Fintech can never fully replace human intelligence, it can undoubtedly strengthen it. Financial companies can leverage the power of Artificial Neural Network or other disruptive tools to construct potent goods and decision-making systems to advance in the area of financial service via the use of computer-based technologies that focus on Big Data analytics [11]. This is leading to significant changes either on an organizational or on individual level. Fintech companies may profit from AI by fulfilling their growth objectives, obtaining a competitive edge, and becoming more responsive to their consumers [22]. Additionally, it can help them save operating costs and streamline internal processes. This will improve users' financial behaviour, which will be advantageous [6]. Fintech applications are creating innovative and engaging methods for consumers to process information [14]. Visualization and data science tools' strength makes it simple to analyze data using apps and turn it into easily understandable insights. Users can make use of complex data to enhance financial decision-making.

Remarkable work by Lagna and Ravishankar [26] aimed at making the world a better place using financial technologies. They concluded with a discussion on how their research will lead to a financially inclusive society. Other work by [11] put forward an adaptive trading model for developing quantitative trading strategies by an intelligent trading system. Their model then intensified to imitation and DRL techniques. The training of trading agents in the real financial market was performed for better counterfeiting. More specifically Le *et al.* [10] wrote a journal discussing the actual situation of the Vietnam Fintech Market with the applications of Machine learning. They made predictions that determine the value for money (VFM) to use Machine learning techniques in their financial institutions in the future. They concluded with the consideration that using machine learning in their financial matters is purely essential because it relates to their existence of them in near future.

In addition to the above work done in the financial markets, other works in stock predictions for finance technology were also performed by researchers. The prediction model is driven by RL in Fintech, according to Shi *et al.* [16], and relies on the heterogeneous information base, including global indicators, tweets, and stock prices. In the RL model, proposals have included the redesigned NDDP for any imputation of data that is missing. Lagna and Ravishankar [26] note that the information trends representing the inherent aspects of monetary markets can be retrieved via LSTM. In the 2 adversarial network models, a heterogeneous fusion of data should depict the market crash or HDFM, thus leaving the world with wonders in the field of finance technology aiming at predictions, management, advising and so on [27].

Some of the literature review discussed above is summarized as shown in the table 1 below:

| Reference | Fintech Area | Methodology | Experimentation |
|---|---|---|---|
| [8] | Trading | Adversarial market model | virtual market setup is showed 4% improvement in portfolio performances |
| [11] | Trading | Intelligent trading system based on an imitative DRL | Trading agent is trained in real financial market and results showed that the model was able to extract robust market features and adaptive in different markets |
| [14] | Smart investment model | RNN based on the genome model | Promising results were shown as Asset Banker RL algorithm took the right investment decisions and made maximum profit |
| [16] | Trading | CNN with Recurrent RL framework for portfolio mgt. | Better results over existing solutions were shown on the cryptocurrency datasets by obtaining higher profits |
| [18] | Finance Portfolio Mgt. | Deep RNNs (GRUs) and DRL for optimizing finance | Authors aimed to present empirical results of their study in the near future |
| [24] | Personalizing debt collections | Aggregated Deep RL and field experiments | Results suggest the use of fewer collection actions by the loan tenets. Also, the significance of personalization in debt collection was evident |
| [25] | For prediction in volatile Stock market | NDDP, LSTM, and adversial learning networks | Performance of prediction was reported better than existing works |
| [27] | Optimizing Trading mgt. system | Q-Learning Algo. and the SARSA Algo. based on RL method | Tested stock prices time series with 500 simulations; both algorithms generally performed well while QLa performed better in individual performances |

Table 1: Summary of some reviewed studies

## 3. Adopted Methodology to Conduct the Study: Using PRISMA guideline

　　Methodology is defined as a system of methods used in a particular area of study or activity while research methodology, presented in the research papers, describes the procedures and tools that can be used to conduct a comprehensive research. As Jain [28] points out, the success of a specific research is determined by the tools and techniques used by the researchers to facilitate completion of a study. Accordingly, the methodology is crucial to inform the intended audience about how the study objectives, research questions, and aims have been investigated throughout the work. Consequently, scholars are inspired to define different approaches, research context, data source, research area, methodology, and strategies to present the whole research methodology comprehensively. Nevertheless, the methodology merely serves the purpose of presenting both experimental as well as computational models that are unique. However, the method being described can be either unique and new, or it can be a bid to offer an improved version of the previously described ones. In this study, we intend to establish the foundation for new research by combining previous groundbreaking and significant researches that have been conducted in the past with this new research. There should be extensive testing performed on the approaches, and ideally, although not always, they should have been applied to demonstrate their value. For readers to use the findings and recommendations of a systematic review, all datasets must be available and understandable to them.

With the aim of exploring developments of RL and its role in FinTech, this study emphasizes on the existing technologies and research on the topic with an explanatory approach. Specifically, the study is secondary research; thus, it is particularly based on desktop research. This means that the study has relied on current and previous studies performed by numerous other scholars in the field of RL and its correlation with FinTech. Given below is the track followed throughout the review:

- Figuring out and defining the aim, design as well as setting of the research study.
- Through previous research, an in-depth description of the main focus of the review which is RL and fintech seperately as well as together.
- A comprehensible and understandable illustration of all the interventions and comparisons between efficiency, accuracy and effectiveness.
- Thoroughly examining the relevance between the previous studies, their successes as well as shortcomings.

As aforementioned, the secondary nature of this study means that the explanatory research method is the most applicable. Generally, exploratory studies denote a research technique exploring the reasons why something happens, especially when limited information is accessible [28]. Exploratory studies can significantly assist to increase understanding of a specific subject, determining why and how a specific phenomenon is taking place, and predicting the future. Such studies are generally concerned with the creation of a knowledge base founded on an intensive search of different information and data [29]. Due to its extensive nature and need to find facts, exploratory studies tend to reveal different outcomes or results that a scholar did not expect. Generally, the findings can contradict the research hypothesis.

The explanatory research method can be viewed as a research that has the starting point with a general idea and uses this research as a mode to spot issues and problems which can serve as a basis for future studies. The main focus here is that the researcher must be open to new contradicting facts and should be capable of moving them accordingly. This is one of the utmost challenges that are a part of this research. With each passing moment, there is a new addition to the previously studied and available literature. Conclusive remarks need to be made accordingly with the literature being revealed. This research was limited due to the limited work available out there, however this can then be served as a grounded theory approach or interpretive research.

The principal purpose of using the exploratory research method in the current survey is the limited resources or studies focused on the actual short- and long-term implications of using RL by financial institutions of FinTech in sustaining and promoting performance. Through the exploratory nature of the study, researchers can get a general concept and apply research as a model of a quick guide to the challenges or concerns that can be discussed in future. Regardless, the foremost task of using the exploratory approach is recognition of reasons for RL being one of the best and most appropriate ways of fostering the performance of financial institutions.

In this line of work to get valuable insights from the studied literature, adopted methodology to conduct the current study incorporates PRISMA technique. The PRISMA technique is an abbreviation for Preferred Reporting Items for Systematic Reviews and Meta-analysis [30]. It ensures improved reporting quality of a systematic review by providing significant clarity in filtering out process of papers. Figure 2 depicts the flow diagram of the current study, carried out using PRISMA.

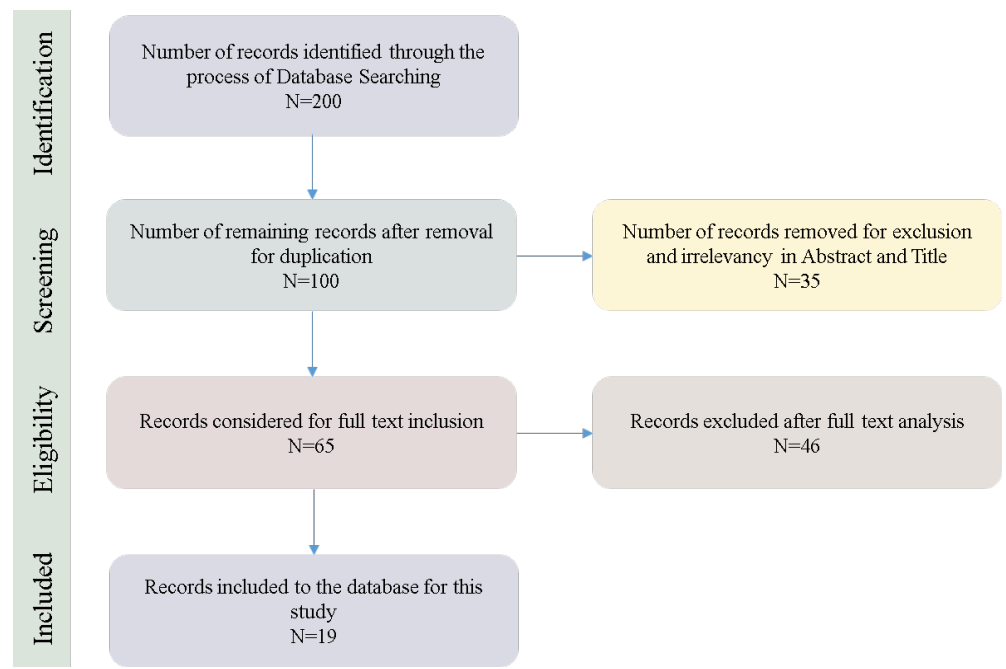

**Figure 2.** PRISMA Flow diagram of our systematic literature review

The methodology of this exploratory systematic research is summarized in the following sections:

### 3.1. Data Collection and Selection

Literature study and data collection was the basic and principal step concerning the topic under review. This step was the most crucial as it served as the base for the whole research and comparisons were made according to the studied literature.

The search procedure was extremely exclusive with the topic being reviewed. The main difficulty was finding true and accurate relevance between our research topic and the available literature for study. We had taken start of our work with the searching process using our research keywords; i.e., RL, Deep Q learning, recurrent unit, recurrent Q network, AI, machine learning and Fintech etc. This way of database search exposed a lot of material for our study. Online research, along with manually available pieces of studies, produced 65 relevant studies. Although more than 200 current studies from 2018 address RL in some way, the rationale for collecting proper and sufficient information on the topic relied upon search terms such as AI and finance, machine learning and finance, RL and finance, and neural networks in finance, among other important key terms. The study focused more on works published in journals such as IEEE, Springer, Procedia Computer Science, Research gate, International Journal of Data Science, Information Systems Research, Journal of Social Commerce and arXiv preprint, among others. After gaining exposure to a lot of related studies, the next task was to screen the work out for duplication. Initially almost 200 research publications were retrieved but after removing duplication, around 100 publications were left. Afterwards, we were engrossed in finding the material exclusive to our research in any way possible i.e., it could be a misleading topic or abstract that made us add it to our work in the first place. Through this filtering, 65 full records were remained to be read fully and understood for being added to our work. After that step, it was found that there were few articles out of those 65 that were not exactly related in a way to be used as a fine material for our survey. Only 19 articles were found relatively closer to our research study.

### 3.2. Data Evaluation

After collecting all the previously available relevant data on the topic being reviewed, the next step was to evaluate the collected articles, papers and research. This step is accessing study inclusion/exclusion details and collecting from study information coding. The 65 publications that showed relevance for the full-text study were deeply studied and evaluated. Focusing recent studies, published works in 2018 and so were considered.We were more focused to work done in the near past so that absolute conclusions and results could be extracted. The studies published in 2018 were 21, 17 were from 2019, 8 from 2020, 13 from 2021, and 6 were considered from 2022. Table 2 shows these figures in a more concise way while Figure 3 presents publications percentages with years.

| Year | No. of Papers | Percentage |
|---|---|---|
| 2018 | 21 | 32.3% |
| 2019 | 17 | 26.2% |
| 2020 | 8 | 12.3% |
| 2021 | 13 | 20% |
| 2022 | 6 | 9.2% |

Table 2: Studies by publication year

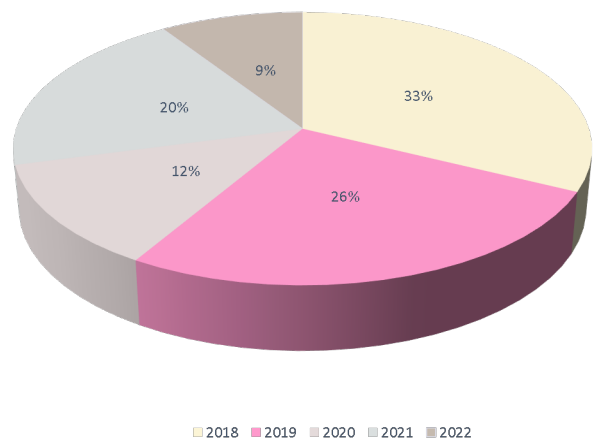

**Figure 3.** Publications percentages with years

Comparing these evaluations, most of the work was done in 2018 and there were fewer works in recent years yet those were powerful to be presented as the core of our research.

*3.3. Data analysis and interpretation*

The most important step was the analysis and interpretation of all the data and gathered studies. This step also included combining the effective results as well as interpreting the analysis results. This collection of results from various sources and research was the way to reach to a concrete conclusion.

*3.4. Results comparison*

Through tables, pie charts, comparison graphs and schematic representations, the results gathered from various studies were combined and compared for a better understanding. These contribute not only to easy access to all the results of different approaches but also made research comprehensible and understandable. The basic idea of review work is mainly to merge various previously adopted approaches and to show them in one representable form, thus this step was performed.

*3.5. Final write-ups and discussion*

Based on result comparisons, further discussions were made on what was retrieved from all of the research work. Exploration of bias and recommendations were made followed by some conclusive remarks.

## 4. Taxonomy

The foremost task of presenting taxonomy of the conducted survey for RL algorithms is the categorization of all RL methods and easy access to the appropriate method for future research. The basic representation of taxonomies is "is-a relationship" whereas for mereologies, it is "has-a relationship" [31]. For this reason, the presented taxonomy is high-level to pave an easy way to reach all the RL methods.

Firstly, more generalized discussion starting from machine learning and its types is made in this section. Afterwards, The study captures notable RL algorithms in Fintech all through.

Following Figure 4 shows an illustration of the types of machine learning.

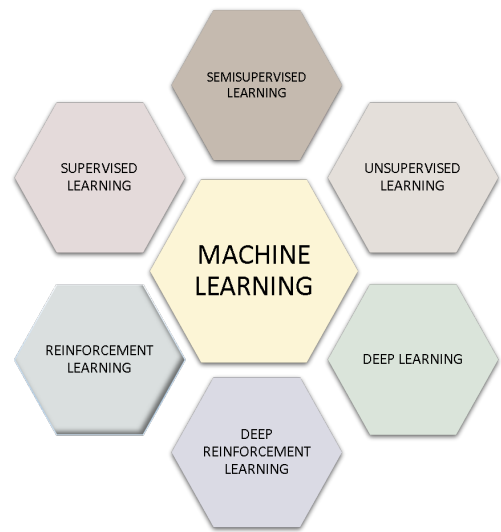

**Figure 4.** Types of Machine Learning

*4.1. Supervised Learning*

It is a type of machine learning with the main objective to train the model on a particular specified set of inputs which results in the establishment of a relationship between the input feature and the prediction outputs [32]. Some of the frequently and customarily used supervised learning algorithms are Decision trees, Naive Bayes, Linear Regression Neural Networks and SVM etc.

*4.2. Unsupervised learning*

In unsupervised learning, the training dataset contains untagged input points data. For obtaining useful and meaningful inferences the algorithm detects patterns [33]. Some widely used algorithms for unsupervised learning can be listed as neural networks for association problems and K-means etc. Dixon *et al.* [34] described that UL is used for drawing inferences from the available dataset. The main goal of UL is to understand and absorb the structure of data rather than predict certain values. UL methods can be categorized as either clustering or factor analyses.

*4.3. Semi-supervised learning*

This type of machine learning, known as semi-supervised learning, lies in the middle of supervised learning and unsupervised learning. Generally in supervised learning, some input points of the data sets are presented as well as the output points corresponding to the given input points. Whereas in unsupervised learning, no specific output value is provided; the underlying structure is tried to infer using the input points. Typically, semi-supervised learning is known for attempting to ameliorate the outcome and performance of these tasks by consuming statistics that are linked to others.

*4.4. Deep Learning*

In deep learning, the main task of an algorithm is to correctly copy the function of the human brain and use it in scientific computing. The most widely used deep learning algorithms are CNNs (Convolution neural networks) as shown in figure 5, recurrent networks and LSTM etc. [35].

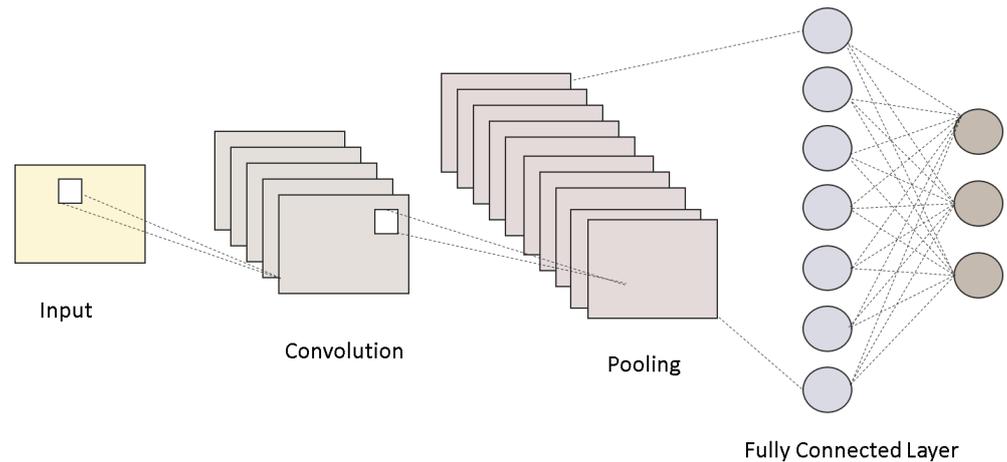

**Figure 5.** Structure of CNNs

*4.5. Reinforcement Learning*

Reinforcement Learning (RL) being a type of machine learning is a learning paradigm that has major concerns with controlling a system so that it can escalate and maximize the performance measures that are used to express a long-term objective. The main discern between supervised and RL is that in RL the feedback given about the learner's prediction to the learner is partial. These predictions are found to have persistent effects which influence the controlled system's future condition. RL is of substantial and great interest due to its practical applications in several domains inclusive of Fintech [36]. General taxonomy of RL algorithms can be visualized in figure 6.

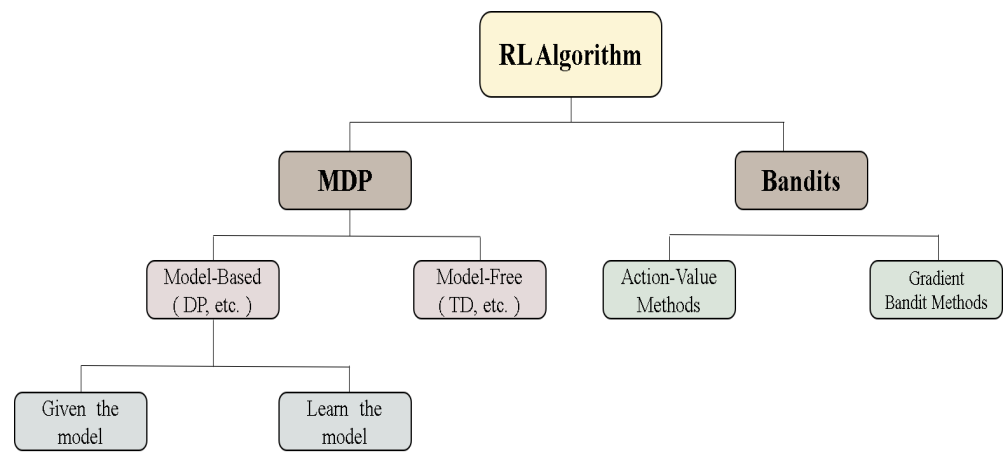

**Figure 6.** General Taxonomy of RL Algorithms

An agent, a policy($\pi$), a reward signal, and a value function are the four primary components that make up a RL system, as shown in Figure 1. In accordance with the provisions of a policy, an agent's behavior at each given instant may be specified. A policy may be thought of as a blueprint that outlines the connections between behavior and perception in a specified environment.
RL is different from supervised learning (SL) as in RL the agent only learns by interacting with the environment repeatedly whereas in SL it can learn in one scan of the complete dataset. The general method is that at the time t an agent performs an action $A_t$ and a reward $R_{(t+1)} = R(S_t, A_t)$ is given to it, after that, the environment is moved on to the next state. The basic and major task for the agent is to learn a way to react to the environment to maximize the total reward as formulated in equation 1, given below;

$$V^\pi(S_t) = \sum_{k=0}^{\infty} \gamma^k R_{t+k+1} \qquad (1)$$

Here, the coefficient $\gamma$ represents the decay factor which usually considers the interest rate in finance [37].

Systems using RL have excelled in a variety of fields including self-driving automobiles, Atari video games, banking, and healthcare [38]. The detailed taxonomy of the RL algorithms noticeably being used in Fintech is given in figure 7.

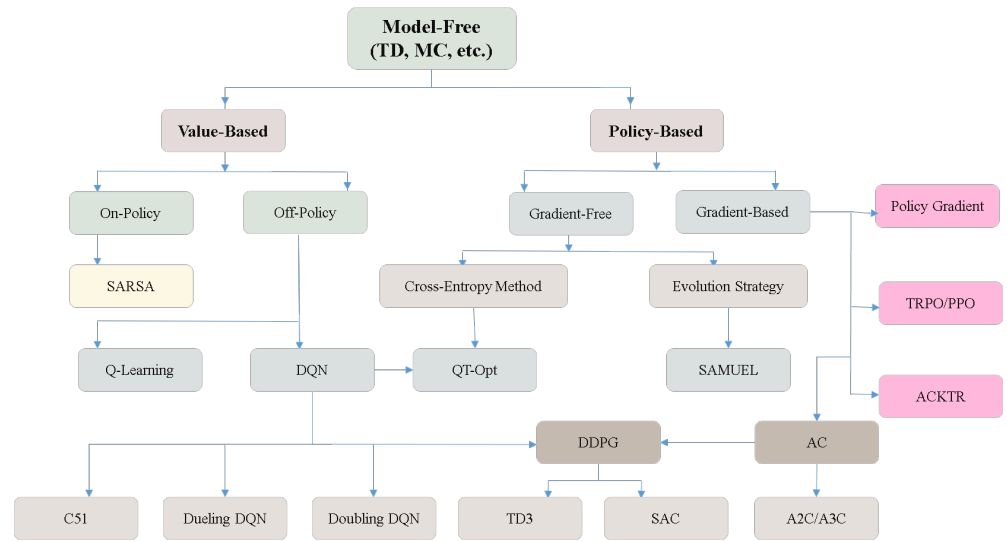

**Figure 7.** Detailed Taxonomy of Model Free RL algorithms notable in Fintech

In figure 6, it can be seen that the RL algorithms are divided into Markov Decision Principle and Bandits. In the MDP, we have further classification into the model-free RL and model-based RL. By calculating the reward and transitional probability function in the MDP context, model-based algorithms sustain an approximation of the MDP model, from which they construct the value function. The value function is then used to derive the policy. Model-free algorithms work by direct learning without inferring the model, a value function or the quintessential policy whereas model-based algorithms do not work that way. They are classified into given the model and learn the model. Model-free algorithms are classified into two main categories as shown in figure 7; these are value-based methods and policy-based methods. Model-free RL in Fintech was seen to be practised in a lot of work, such as Sato [39] used it for portfolio optimization. Model-free RL is further classified into Value-based methods and Policy-based methods, as discussed below:

4.5.1. Value-Based Methods

Here, the method of the traditional value-based algorithm is presented in an infinite time horizon manner with discounting action spaces, finite state and stationary policies. Value-based methods are further divided into on-policy and off-policy methods.

**On-policy methods:** These approaches focus on studying and improving the policies that underlie decision-making, thus form the foundation of decision-making. These techniques seek and make an effort to first provide an estimation, and then to alter the policy that is currently being used for decision-making.

Here in On-policy methods, the policy that is being worked on is usually soft by soft it means;

$$\pi(a|s) > 0 \; for \; all \; s \in S, a \in A \tag{2}$$

One of the examples of these policies can be seen as $\varepsilon$-greedy policies by this we mean that in most cases they choose those actions which have a maximal estimated action value but when they have a probability $\varepsilon$ a random action is selected instead of those actions which have a maximal estimated action value [40].

One of the most common approaches of On-policy RL, known as SARSA (which stands for state-action-reward-state-action), evaluates the value of the policy that is now being followed. During this phase of the algorithm, the agent analyzes the situation and determines the best course of action to take. There is no difference between the policy that is used for acting and the policy that is used for updating.

Update rule is used to define a state action and reward state action. The update rule can be seen as;

$$Q(S_t, A_t) \leftarrow Q(S_t, A_t) + \alpha [R_{t+1} + \gamma Q(S_{t+1}, A_{t+1}) - Q(S_t, A_t)] \tag{3}$$

This is done from a $S_t$, nonterminal state, after every transition. $Q(S_{t+1}, A_{t+1})$ is set to 0 if is the terminal point. Every element of the quintuple i.e, $S_t, A_t, R_{t+1}, S_{t+1}, A_{t+1}$ which is utilized in the making of the switching from one to the other state action pair used by the rule stated above. This gave rise to the algorithm named SARSA. The factor $q_\pi$ is being continually estimated for the policy $\pi$ and $\pi$ is being changed at the same time towards the greediness concerning $q_\pi$. When you want to maximize the benefits of an agent that is exploring its environment, using on-policy RL may be quite helpful.

**Off-policy methods:** In contrast, an off-policy is not affected by the actions of the agent in any way. It analyzes the situation and determines the best course of action, irrespective of the agent's goals. The evaluation of the rewards based on these approaches does not take into account the actions that are now being taken. They take into consideration all of the alternative actions that may be taken at the moment in order to maximize the benefit that can be achieved from the next step[41]. The main off-policy method includes Q-learning and Double Q-learning algorithms.

*Q-learning:* This algorithm memorizes the value of an action in a particular state. It is model-free; hence, it doesnâĂŹt need an environment model. Inclusively, it can take care of stochastic transition issues and rewards despite requiring adaptations.

The Q-learning algorithm is defined by the update rule given below:

$$Q^{new}(S_t, A_t) \leftarrow Q(S_t, A_t) \alpha [R_{t+1} + \gamma \max_a Q(S_{t+1}, a) - Q(S_t, A_t)] \tag{4}$$

Here, $\alpha$ is the learning rate that has values $0 < \alpha \leq 1$ and $R_{(t+1)}$ is the reward being gained here when moving to the state $S_{(t+1)}$ from the state $S_t$ and $\gamma$ is the discount factor. On the other hand, a point that can be seen here is that the new state $Q^{new}(S_t, A_t)$ is explained below;

1. The newest value being weighed by the learning rate $(1 - \alpha)Q(S_t, A_t)$. A fact is that the learning rate values closer to 1 make rapid and fast changes in the Q function.
2. The second factor is $\alpha R_{t+1}$
3. The third factor that comprises the new state is $\gamma \max_a Q(S_{t+1}, a)$: this is the maximum reward, weighted by the discount factor and the learning rate that can be achieved from the state $S_{(t+1)}$.

Here, the optimal function values are directly approximated by the function Q liberated of the policy that is stuck upon for this reason it constitutes an off-policy algorithm [42]. Since, the policy chooses which state-action pairs are reached and changed, it still has an impact. However for proper convergence, it is necessary that all pairings continue updating. Obtaining the optimal state value function (its convergence in the deterministic and stochastic settings is ensured by Q-Learning) determines the action of the agent rather than finding the optimal policy.

With transaction costs taken into account during each re-balancing period, [43] explained that by using Q-Learning (without a neural network) in discrete-time market conditions; one may maximize a portfolio that comprises a hazard-free asset (cash) and a risky asset (stock market portfolio).

The adaptive market hypothesis's prediction that the market would not be able to accept new details as rapidly as it arrived led to the development of the Q-learning and SARSA methodologies. For financial trading, [27] also employed Q learning and SARSA and also compared both.

*Double Q-learning:* The fact that the identical Q function, as in the current action selection policy, is utilized for the calculation of future maximum approximated action in Q-learning, it can often hyperbolize the action values in noisy environments which results in slowing the learning [44]. To solve this problem, another off-policy technique was introduced called Double Q-learning. In this technique, a different policy is used for the evaluation of the next value than that which is used for the evaluation of the next action.

Here in this technique, two separate value functions are trained using separate experiences as shown below in a mutually symmetric pattern through equation 5 and equation 6:

$$Q^A_{t+1}(S_t, A_t) = Q^A_t(S_t, A_t) + \alpha_t(S_t, A_t)(r_t + \gamma Q^B_t(S_{t+1}), arg \max_a Q^A_t(S_{t+1}, a) - Q^A_t(S_t, a_t) \quad (5)$$

and

$$Q^B_{t+1}(S_t, A_t) = Q^B_t(S_t, A_t) + \alpha_t(S_t, A_t)(r_t + \gamma Q^A_t(S_{t+1}), arg \max_a Q^B_t(S_{t+1}, a) - Q^B_t(S_t, a_t) \quad (6)$$

*Deep Q Neural Network (DQN)*

The DQN method builds a matrix to help the agent that is working to choose the precise course of action that would maximize its reward in the future. Characterizing a Q-table, however, becomes incredibly difficult and time-taking as the number of environmental states and activities rises. The state is provided as the input for deep Q-learning, and the Q-value for each action is created as the output. The next course of action is determined by the Q-maximum network's output when the Epsilon-Greedy Exploration method is used. The Bellman equation is used to update the network weights since the goal value is uncertain.

These value-based methods were frequently used in finance technology whether be stock trading, portfolio management or cryptocurrencies etc.

4.5.2. Policy-Based Methods

Here, our target is model-free policy-based methods where a parametrized policy is trained without inferring the value function. Policy-based methods are further divided into Gradient-free and Gradient-Based methods. The methods that were seen to be used in finance technology were mainly gradient-based. Actor-Critical Approaches, Trust Region Policy Optimization (TRPO), and Proximal Policy Optimization (PPO), Soft Actor-Critic (SAC), among others, are examples of policy-based methods.

**Proximal Policy Optimization (PPO):** The PPO algorithm uses a policy gradient approach for RL. The goal was to create an algorithm that, although utilizing solely first-order optimization, had the data efficiency and dependable performance of TRPO [45]. This algorithm updates policies using the below equation:

$$\theta_{k+1} = arg \max_\theta \underset{s,a \sim \pi_{\theta_k}}{E} [L(s, a, \theta_k, \theta)] \quad (7)$$

Here, $L$ can be presented as;

$$L(s, a, \theta_k, \theta) = min\left(\frac{\pi_\theta(a|s)}{\pi_{\theta_k}(a|s)} A^{\pi_{\theta_k}}(s, a), clip\left(\frac{\pi_\theta(a|s)}{\pi_{\theta_k}(a|s)}, 1 - \epsilon, 1 + \epsilon\right) A^{\pi_{\theta_k}}(s, a)\right) \quad (8)$$

Where, $\epsilon$ is a (small) hyperparameter that generally indicates how far the new policy can deviate from the previous one.

**Soft Actor-Critic:** The Soft Actor-Critic (SAC) technique tides over the gap between stochastic policy optimization and DDPG-style methods by optimizing a stochastic policy in an off-policy manner. Although, it was released roughly concurrently with TD3; it is not a straight replacement for TD3. Nevertheless, it features the clipped double-Q technique and gains from target policy smoothing because of the policy's intrinsic stochasticity.

A measure of the randomness of a random variable is called entropy. Let $P$ be the probability mass of the random variable $x$:

$$H(P) = \mathop{\mathbb{E}}_{x \sim P}[-logP(x)] \qquad (9)$$

Each time step in entropy-regularized RL includes an additional reward for the agent. This transforms the RL issue into:

$$\pi^* = arg\max_{\pi} \mathop{\mathbb{E}}_{\tau \sim \pi} \sum_{t=0}^{\infty} \gamma^t (R(s_t, a_t, s_{t+1}) + \alpha H(\pi(.|s_t))) \qquad (10)$$

The trade-off coefficient is $\alpha > 0$.

*4.6. Deep Reinforcement learning*

Deep Reinforcement learning (DRL) is an amalgamation of the deep learning method and RL method. DRL has proved to resolve many complex decision-making and prediction issues that were unsolvable before [46]. It has made its mark in the fields of finance technology to a great extent. Many works are available in the context of DRL and finance technology. The figure 8 below shows the basic agent environment interaction in DRL.

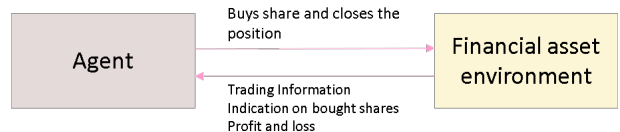

**Figure 8.** Interaction between Agent and environment in DRL

## 5. Review

Reinforcement learning has contributed to optimal execution, Robo-advising (Wealth Management), portfolio management, cryptocurrencies and trading etc.
Through the lens of supervised and unsupervised algorithms, RL has participated to the rise of trading bots in the financial or stock market. Bots characterized by RL architectures can sufficiently learn from the stock and trading market environments while also interacting effectively with them [11]. Specifically, the bots in the trading arena rely on a trial and error approach to optimize their learning strategy.
The second use case of RL in Fintech is its use in the development of chatbots [25]. Analysis indicates that the application of automated chats in the field of finance has produced innumerable benefits to the trade. Specifically, chatbots can effectively serve as brokers while also providing real-time quotes to the user operators [16]. Secondly, conversational user-interface-based chatbots have adequately assisted customers to resolve complex and demanding issues rather than relying on company staff or backend support. RL in conversational user interfaces has greatly helped multiple firms to save time and relieve the support team from endless and repeatable tasks.
In Fintech, RL has also contributed greatly to the optimization of risks Wang *et al.* [47], particularly in peer-to-peer (P2P) lending; P2P lending denotes a means of offering businesses and individuals loans via online channels [25]. The online services perform the task of matching lenders with investors. In the existing marketplaces, RL is important because it can effectively appraise the scores of borrowers credit to limit lending risks [8]. Additionally, RL in risk optimization can sufficiently predict yearly returns. The rationale for this is that because online commerce is linked with low overheads, lenders can generally expect better returns than investment products and savings given by financial banks.

**Reward functions** were used in the field of RL for fintech to improve the predictability level. As the agent's behaviour is controlled by the reward signal, it is an integral part of the RL paradigm. The "ought" behaviour of the agent is described by reward functions. To put it differently, they contain "normative" material that specifies what you want the agent to do. Therefore, you must make it up to the degree that the reward function establishes the agent's motives. There are no formal restrictions, however, the agent will learn more efficiently if your reward function is "better behaved."

*5.1. RL in Goal-based Wealth Management*

Practices have been made on the portfolio optimization problems using the framework of Goal-based wealth management. In contrast to the standard deviation of the portfolios, the risk is defined in this formulation as the likelihood that investors will not achieve their objectives after a while. However, as demonstrated, there is a mathematical relationship between the original mean-variance problem and the GBWM problem. While this problem is static, [48] employed DP to solve the long-horizon portfolio problem to solve the dynamic version of the GBWM problem. Similarly, [49] employed the Q learning algorithm and presented some experimental results. There were some inputs to the problem the covariance matrix of returns which were later used for the computations of 15 portfolios. The outputs were the algorithm used, training epochs and the final value of the functional outcome.

The process encompasses Q learning which was chosen based on the fact that it was using the epsilon greedy algorithm for the choosing action. After that Dasa and Varmaa [49] verified the results of the Q learning algorithm by comparing and contrasting the value function given by V(W(0),0) at t=0. Furthermore, the $\gamma$ parameter was used as a discount factor in the Q-learning algorithm with the aim of future rewards. The fact that no discounting is needed for the rewards used in the GBWM problem forming $\gamma$ was set to 1. The window of Q values that were averaged collectively was managed by the parameter. Experimentally, they found that the value of = 0.1, which corresponds to a moving average over the most recent 10 Q values, performed quite well. This value was shown by plotting the moving average squared difference between the Q-tensors from different epochs. The algorithm stabilized after 20,000 epochs. There was a final expected reward function defined as:

$$[W(T-1), T-1]$$

This is known as the value function which is used for each next node and it gave optimal action and expected terminal reward.

*5.2. RL in Retirement plan optimization*

Dixon and Halperin [50] discussed different RL algorithms for wealth management and Robo-advising such as G learner and GIRL algorithms. G learning is a RL method with a stochastic policy; it can be seen as an entropy regularized Q learning that is more fitting while working with turbulent data. G learners amount to a generative RL model. In this study, they observed the reward functions of various techniques and reduced them to one non-linear equation. The final obtained equation was;

$$G_t^\pi(x,a) = \hat{R}(x_t, a_t) + \mathbb{E}_{t,a}\left[\frac{\gamma}{\beta} \log \sum_{a_{t+1}} \pi_0(a_{t+1}|x_{t+1}) e^{\beta G_{t+1}^\pi(x_{t+1}, a_{t+1})}\right] \quad (11)$$

The term "G-learning" refers to an off-policy time-difference (TD) technique that generalizes Q-learning to noisy settings where an entropy-based regularization is acceptable. It may be used to express G-learning for finite values in a RL scenario with observed rewards.

The experimentation of solving the optimization problem was made by the researcher. They ended with the results and findings that both the discussed algorithm can either be used separately or collectively. In particular, by modelling the real human agents as G-learners and then using GIRL to deduce the latent aims of these G-leaners, their amalgamation might be employed in Robo-advising. After successfully simulating the top human investors, GIRL will be able to provide clients with a Robo-advising service that will allow them to outperform the best investors overall.

*5.3. RL and cryptocurrency price prediction*

Cryptocurrencies have earned the title of the famous key factor in the business as well as financial opportunities during recent advancements. The major activity here is that with regards to the volatility of high prices and inconsistency in the market the cryptocurrency investment is not visible.

Based on problems with the vivid previous approaches to price prediction, [51] presented a machine-learning (RL) technique for price prediction for some financial institutions. They presented an outline of the price prediction process which can be seen in Figure **??**.

The system proposed by [51] constitutes a RL algorithm for the prediction and analysis of the price as well as their work also ensures a secure transaction environment via a blockchain framework. The result of the work showed that the accuracy gained through the RL algorithm in the price prediction was even better than other state of art algorithms.

Trading cryptocurrencies has recently grown in importance and popularity [52]. Research revealed that there are three important data sources contained in the cryptocurrency price prediction. A market statistic makes up the first one. The second is blockchain network information, which includes hash rate, transaction count, and fee information. Google tweet and trend package are the final two [53]. It is to be observed that whatever actions we take whether good or bad if the reward function gains high rewards then that action was good.

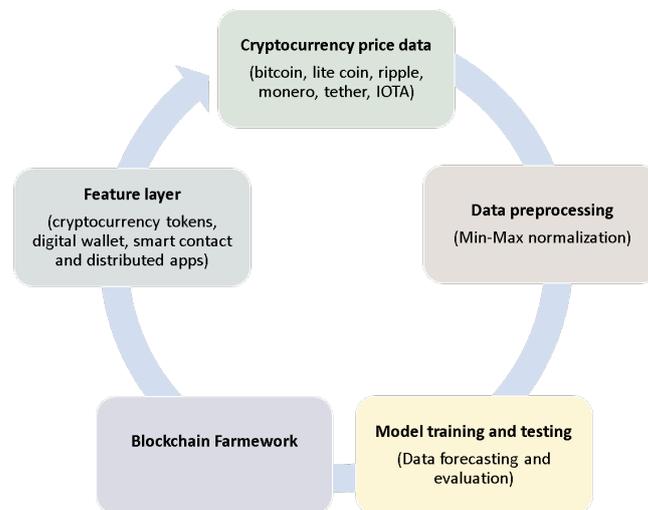

**Figure 9.** Price prediction process outline

The work done contributed to the following;

1. Application of RL for the prognosis of the prices of Litecoin and Monero coins.
2. Evaluation of the performance of the system using the matrices e.g. Mean Absolute error (MAE), Mean absolute percentage error (MAPE), Mean squared error (MSE) and Root mean squared error (RMSE).
3. Utilization of framework of blockchain for the secure and certain environment for predicting prices.

The training and testing dataset for Litecoin and Monero was divided 80% and 20% respectively. Results were evaluated for the normal and their proposed model. The results were captured with time frames of 3, 7 and 30 days and results were gained for the values of the coins for MAE, MAPE, MSE and RMSE. Tables 3, 4, 5 and 6 show the results presented by the work of [51].

| Model | Coin Type | Mean Squared Error(MSE) | | |
|---|---|---|---|---|
| | | 3 days | 7 days | 1 month |
| Normal | Litecoin | 196.5063 | 28.2988 | 287.2785 |
| | Monero | 232.0476 | 31.9216 | 524.8219 |
| Proposed | Litecoin | 6.3949 | 5.2429 | 21.8329 |
| | Monero | 11.8142 | 31.3669 | 410.9197 |

Table 3: MSE values for 3, 7 and 30 days prediction

| Model | Coin Type | Root Mean Squared Error(RMSE) | | |
|---|---|---|---|---|
| | | 3 days | 7 days | 1 month |
| Normal | Litecoin | 14.0572 | 6.3255 | 17.9273 |
| | Monero | 16.1076 | 6.6618 | 23.9958 |
| Proposed | Litecoin | 3.3097 | 3.1438 | 5.6632 |
| | Monero | 4.3826 | 6.6116 | 21.3548 |

Table 4: RMSE values for 3, 7 and 30 days prediction

| Model | Coin Type | Root Mean Squared Error(RMSE) | | |
|---|---|---|---|---|
| | | 3 days | 7 days | 1 month |
| Normal | Litecoin | 14.0572 | 5.3749 | 15.8082 |
| | Monero | 16.1076 | 4.9481 | 20.7854 |
| Proposed | Litecoin | 3.3097 | 2.6536 | 4.9246 |
| | Monero | 4.3826 | 5.8589 | 20.6614 |

Table 5: MAE values for 3, 7 and 30 days prediction

| Model | Coin Type | Root Mean Squared Error(RMSE) | | |
|---|---|---|---|---|
| | | 3 days | 7 days | 1 month |
| Normal | Litecoin | 22.5067 | 7.4219 | 16.9552 |
| | Monero | 23.3272 | 6.5516 | 20.2365 |
| Proposed | Litecoin | 4.0048 | 3.1692 | 5.9518 |
| | Monero | 5.1838 | 7.3865 | 20.4594 |

Table 6: MAPE values for 3, 7 and 30 days prediction

### 5.4. Contributions of RL to trading in the stock market

The analyzed studies have also significantly contributed to the effectiveness of the overall stock-trading environment in the financial markets. [16] found that RL has greatly contributed to setting effective price strategies. Dynamic and complex changes in stock price emerge as some of the major challenges in aligning and understanding stock prices. Accordingly, understanding the underlying concepts requires the use of RL models or algorithms such as the gates Recurrent Unit (GRU) networks. The algorithm or network can perform well based on RL.

The survey findings have also highlighted the powers of RL in the field of finance by describing its actual correlation with the maximization of profits using minimal investments in the capital [8]. Combining all the principles of RL, portfolio management, recommendation systems, strategic setting of prices, and risk optimization means that RL has greatly contributed to the field of finance in diverse ways [11]. In capital management, for instance, strategic planning and automation are the ways that the use of RL technology has fostered the welfare of trading and the stock market not only in the United States but also globally.

A special data augmentation technique is introduced by [54] through which the data set would be increased by an adequate amount and more precise and correct training would be performed which would contribute to satisfactory results. They further chose the method based on skewness and kurtosis to choose stocks that were to be traded with the usage of the presented algorithm. The experimentation showed that the model proposed by them gave decent returns as results compare to the buy and hold strategy. Figure 10 presents the agent environment interaction.

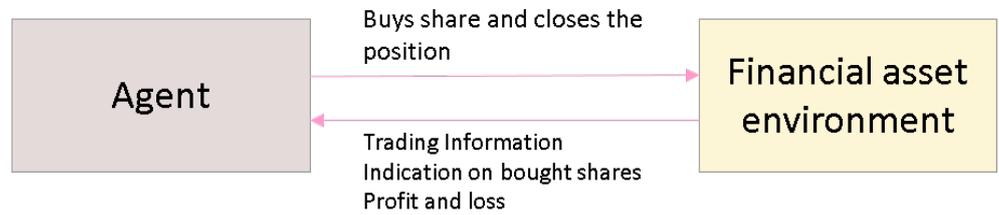

**Figure 10.** Interaction between environment and agent

**Rewards:** The so-called returns, rather than the actual prices, are typically examined to understand the evolution of stock market prices and create lucrative portfolios. Returns, which represent the relative change in price over each time step, are a more reliable method than utilizing actual prices since they have undergone normalization, which improves generalization across various stocks.

The experimentation continued in a way that they selected 3 stocks from the 11 stocks. These three stocks were YZGF (603886. SH), KLY (002821. SZ), and NDSD (300750. SZ) from the Tushar database. Annualized return (AR) and Sharp ratio (SR) were used by them as the gauging metrics and the results were also obtained accordingly. The final portfolio value indicates the final trading stage portfolio value. The direct return of the portfolio every year is shown by the annualized return. The annualized standard error demonstrates our model's stability. The evaluation is made using the Sharpe ratio, which combines return and risk [55].

The benefit of Annualized return usage was its capability to put various periods into the same scale which was useful in comparing different stock returns using different agents. On the other hand, the Sharp ratio can be referred to as a risk-adjusted return which uses the return $R_p$, the standard deviation and risk-free rate $R_f$ of excess return $\sigma_p$. The formula is:

$$SR = R_P - R_f \sigma_p$$

The RL algorithms that were used were the DQN, Proximal Policy Optimization and Soft Actor-Critic.

Buy and hold was the baseline set for the trading strategy. It is used for the securities that are being held for quite a long period. If you purchase and hold, it can be because you think the short-term volatility that comes with the stock investment will be worth it for the long-term profits [56]. It is a passive investing style which can be simply summarized in a way that B&H strategy users believe in the importance of time in the market than timing the market. That being said holding on to stock is easy than timing the market perfectly.

For comparing with B&H they set 0 as the risk-free rate. The sharp ratio is a relative metric which was implicated in comparing the risk-adjusted returns of various trading plans.

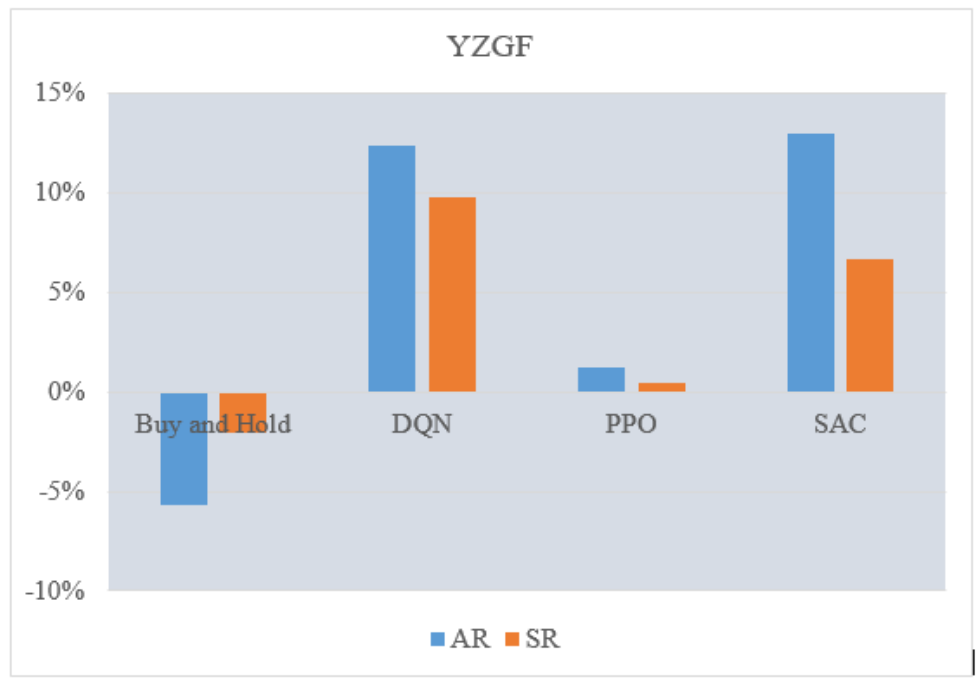

**Figure 11.** Comparison results of YZGF test set

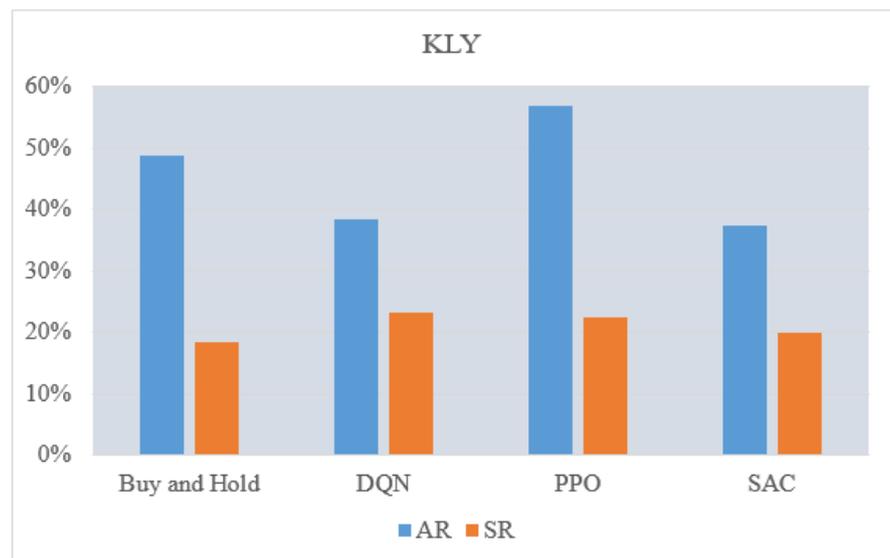

**Figure 12.** Comparison results of KLY test set

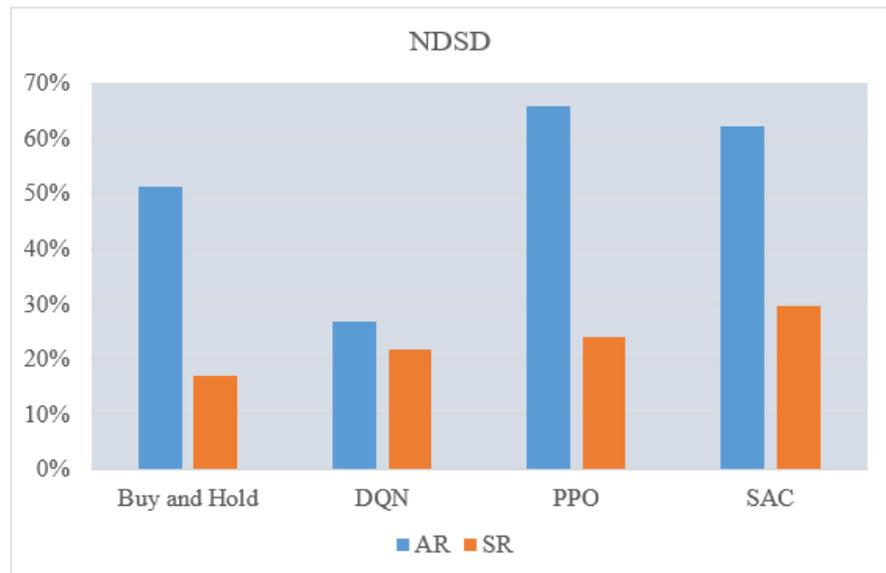

**Figure 13.** Comparison results of NDSD test set

Results of (Yuan, Wen and Yang, 2020) can be seen in figures 11, 12 and 13 which showed that Proximal policy optimization was more stable and accurate than the other methods. The main reason for this conclusion was that it's SR and AR were outperforming the buy and hold strategy continuously. In contrast to training results, DQN and SAC consistently outperform the B&H method in SR and frequently outperform it in AR.

Additionally, DQN or SAC occasionally achieve the highest SR of all strategies or the highest risk-adjusted return. It is important to note that their training dataset's stock and period are completely different from those of the testing dataset. It can therefore be concluded that the agents' excess return can be attributable to the market law they discovered on their own where neither time nor stock affects the law. In experimentation, PPO is a steady algorithm; nonetheless, DQN and SAC occasionally get greater AR and SR.

A study performed by [57] spread light on the topic of RL in stock trading. Their work studied not only studied the usage of RL algorithms in stock trading but also evaluated the approach to real-world stock data. They compared their work algorithm (RL techniques) with other techniques utilized in the world of stock prediction. They believed that in a market where each next step is unpredicted, their work based on RL will make nearly accurate decisions for stock trading.

Experimentation began with them evaluating three variants of Deep Q-learning which were vanilla Q-learning [58], Double DQN [44,59] and Dueling Double DQN [60]. Their data set included daily stock prices for more the seven thousand US-based stocks that were collected up to 10 November 2017. The testing period for each stock was set from 1st January 2017 to 10th November 2017 whereas the training time was from 1st January 2015 to 31st December 2016. This gave them a total of 504 training and 218 testing samples.

Results of the work revealed that DQN compared to Dueling DQN and Double DQN yielded the highest average profit. The results obtained were consistent with the other studies too [61]. DQN produced larger volatility than the other two approaches, as expected. They showed the breakdown of profits produced by each model. It was obvious that Double Q-Network was a business, hence occasionally it made a loss.

In the paper [57], they examined how DQN was used in stock trading. They tested Deep Q-effectiveness Network's using sizable real-world datasets. With DQN, they can trade stocks without extra optimization steps like with other supervised learning approaches. RL algorithm variations based on Q-learning can produce strategies that, on average, make a profit using only a small number of samples.

*5.5. Effectiveness of RL in portfolio management*

Portfolio management is also another major theme emerging in the connection between RL and Fintech. From the analyzed literature, [11] offer the best argument about the actual implication of DRL and portfolio management. The findings indicate that with the assistance of Deep Policy Network RL, possibilities of asset optimization allocation can be achieved over time. RL, in this regard, can significantly assist in achieving three major goals. The first goal is that RL can promote the success and efficiency levels of human managers. The second goal is that RL can lessen organizational vulnerabilities. The third objective is that RL can foster the levels of return on investments (ROIs) regarding organizational profits.

An asset allocation method termed meta policy was introduced by Jangmin O et al. in 2006 to dynamically adjust the asset allocation in a portfolio to optimise returns [62]. A general illustration of a portfolio management system can be seen in figure 14.

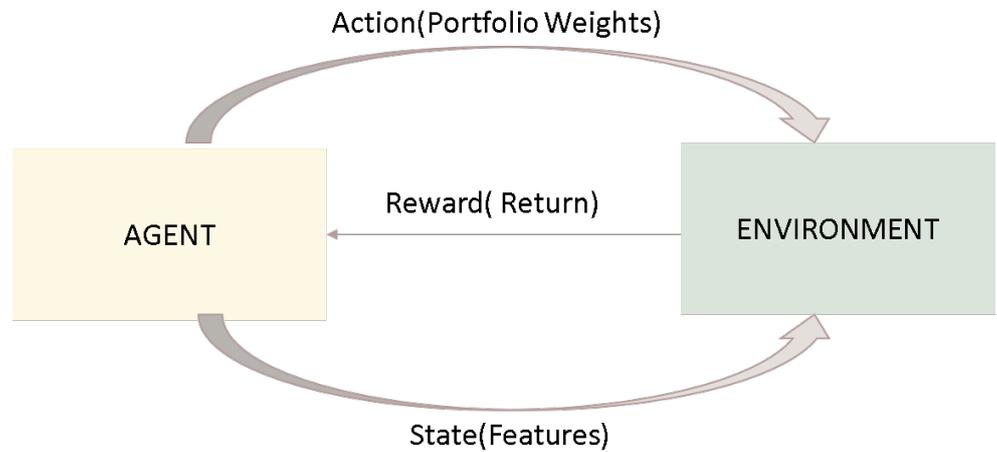

**Figure 14.** A general illustration of the Portfolio Management System

The financial world has recently noticed a rise in the popularity of RL models [35]. The cardinality constraint, floor and ceiling constraints, round-lot restriction, pre-assignment constraint, and class constraint are a few of the real-world limitations that practitioners of portfolio optimization take into account. [63].

In their study from 2020, Wang et al made the first effort to leverage an algorithmic trading-related concept. His development of an HRPM is a specific example. Two decision processes were arranged in a hierarchy in their system. While the low-level policy chooses at what price and what quantity to place the bid or ask orders within a condensed time window to accomplish the high-level policy's objectives, the high-level policy modifies portfolio weights at a lower frequency.

To address portfolio optimization issues, both value-based techniques (Q-learning, SARSA, and DQN) and policy-based algorithms (DPG and DDPG ) have been used. The state variables are repeatedly made up of duration, property prices, property former returns, current asset holdings, and the balance that is currently available.

For value-based algorithms, [43] considered the portfolio optimization problems of a risky asset and a risk-free asset. The Sharpe ratio, differential Sharpe ratio, and profit were three different value functions that were used to assess the performance of the Q-learning Recurrent RL (RRL) algorithm.

The reward $r(s, a, s^{'})$ was calculated that is the change in the portfolio value when at a state $s$ and action $a$ is taken that makes it arrive at a new state $s^{'}$. Figure 15 shows recurrent RL in action.

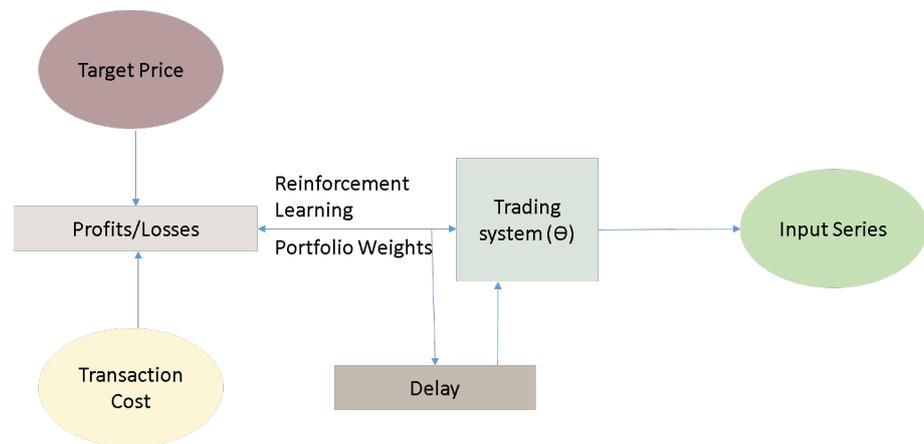

**Figure 15.** Illustration of Recurrent reinforcement learning in portfolio management

The last action is used as an input by the policy-based RRL algorithm. They concluded that the Q-learning algorithm performs less consistently than the RRL method and is more susceptible to the selection of the value function. Furthermore, it is concluded that the profit should not be the preferred reward function, but rather the (differential) Sharpe ratio.

## 6. Results And Discussion

This survey discusses and summarizes different aspects and facets on the role of RL in Fintech including complexity, scalability, accuracy, profitability, risk management and speed etc. PRISMA guideline for reporting systematic reviews is utilized to conduct this study which help to present more complete reporting of systematic reviews. Among the 65 studies selected for the current research survey, only 19 studies accurately focused on the interconnection between RL and FinTech. Using PRISMA approach, 19 shreds of literature are chosen for comprehensive analysis, addressing the issues, merits, demerits, and future of RL in the finance or monetary market. The results revealed that the study on the challenges and potentials of RL in Fintech has not been fully explored and researched comprehensively.

To be focused on the research purpose, along with the most appropriate studies associated with the study topic, 13 specific studies contributed significantly in understanding the connection between RL and the Fintech sector. The rationale for considering these thirteen sources as appropriate to the current study objectives is that they satisfy all the keyword and scholarly study factors required to any research survey. Table 7 shows the details of these studies, meeting the requirements of the survey paper.

Our work indicated that RL in the fields of Fintech was useful compared to other classical and traditional methods. RL stock trading techniques were mostly Model-Free, specifically Value- and Policy-based. The results achieved more accuracy than those obtained using conventional fintech strategies such as Buy and Hold. A little work was done concerning RL-based techniques in cryptocurrency predictions and goal-based wealth management systems. Most of the work was observed in Portfolio optimization using RL and gave significant results in contrast to other algorithms.

In addition, reinforcement learning was able to simulate more efficient models with more realistic market constraints, yet there were limitations in terms of scalability of the model. DRL further addresses the RL algorithms' scalability issue, which is essential for rapid market and user development and logically works with greatly wanted high-dimensional settings in the financial market.

Reinforcement learning has much potential, but it may be challenging to apply and only has a few uses. Deployment issues arise from this type of machine learning's dependence on environment investigation. For instance, a robot that uses reinforcement learning to navigate a complex physical environment would look for new states and develop new behaviors. However, it might not be easy to continuously make the proper choices considering how quickly the atmosphere shifts in the real world. Due to the time needed to guarantee that the learning is carried out correctly, the method's

Table 7: Studies meeting the survey paper requirements

| Author(s) | Journal Title | Publication |
|---|---|---|
| Kuo *et al.* [8] | Improving generalization in reinforcement learning–based trading by using a generative adversarial market model | IEEE |
| Liu *et al.* [11] | Adaptive quantitative trading: An imitative deep reinforcement learning approach | Proceedings of the AAAI conference on artificial intelligence |
| Shi *et al.* [16] | GPM: A graph convolutional network based reinforcement learning framework for portfolio management | Elsevier |
| Hu and Lin [18] | Deep Reinforcement Learning for Optimizing Finance Portfolio Management | IEEE |
| Yang *et al.* [24] | Personalizing debt collections: Combining reinforcement learning and field experiment. | 1st International Conference on Information Systems (ICIS 2020) |
| Khuwaja *et al.* [25] | Adversarial Learning Networks for FinTech applications using Heterogeneous Data Sources | IEEE |
| Dixon and Halperin [50] | G-learner and girl: Goal based wealth management with reinforcement learning. | ArXiv Preprint ArXiv:2002.10990. |
| Shahbazi and Byun [51] | Improving the cryptocurrency price prediction performance based on reinforcement learning. | IEEE |
| Yuan *et al.* [54] | Personalizing debt collections: Combining reinforcement learning and field experiment | 41st International Conference on Information Systems (ICIS 2020): Making Digital Inclusive: Blending the Local and the Global |
| Dang [57] | Reinforcement Learning in Stock Trading | Springer |
| Le *et al.* [64] | Applications of Machine Learning (ML)-The actual situation of the Vietnam Fintech Market | Journal of Social Commerce |
| Wang *et al.* [65] | Deep Stock Trading: A Hierarchical Reinforcement Learning Framework for Portfolio Optimization and Order Execution. | arXiv preprint arXiv:2012.12620. |
| Das and Varma [66] | Dynamic Goals-Based Wealth Management Using Reinforcement Learning | Journal Of Investment Management (JOIM) |

insignificance and demand on computer resources may be constrained.

Another challenge reinforcement learning faces in fostering financial decision-making and operations is the limited samples of data available to make appropriate policies, strategies, and decisions. Kuo *et al.* [8] refer to sample efficiency as an algorithm retrieving the most resources from a specific sample. Furthermore, sample efficiency denotes the experience level that an algorithm should produce in a training session to obtain efficient performance. The major challenge or issue is that reinforcement learning is assumed to take considerable time to reach efficiency. Liu *et al.* [11] contends that because the action space and state space could be unprecedentedly huge, it is not feasible to ask for a sample size that exceeds the fundamental thresholds established by the ambient tabular settings specific to a dimension.

## 7. Conclusion

Although reinforcement learning has captivated a lot of consciousness in the field of AI, there are still many limitations to its adoption and use in the actual world. Despite this, several research articles on its theoretical applications present some practical use cases. As long as a distinct reward is available, reinforcement learning can be applied to a circumstance. Reinforcement learning algorithms in FinTech can distribute scarce resources across various tasks as long as a broad objective is pursued. Saving time or preserving resources would be an objective in this situation. Reinforcement learning has been used in a few restricted experiments in robotics. Robots using this type of machine learning may be able to learn skills that a human supervisor cannot demonstrate, apply previously taught skills to new tasks, or optimize even without an analytical formulation. The financial sector aims to continue getting benefits significantly from the reinforcement learning technology. The findings of the conducted study, based on the PRISMA technique which aimed to provide complete reporting of systematic reviews, indicate that the reinforcement learning in Fintech has allowed financial companies to optimize their portfolio, reduce credit risk, effectively manage a portfolio, set effective price strategies, and reduce capital. Despite its strengths, reinforcement learning has always been overlooked. However, the current study shows obvious merits of reinforcement learning by highlights some of its best use cases.